# IDEAS OF PHYSICAL FORCES AND DIFFERENTIAL CALCULUS IN ANCIENT INDIA

T.E.Girish   and C.Radhakrishnan Nair*
Department of Physics,
University College,
Trivandrum 695 034 ,INDIA
( * Retired  Faculty)
Email: tegirish5@yahoo.co.in

## Abstract

We have studied the context and development of the ideas of physical forces and differential calculus in ancient India by studying relevant literature related to both astrology and astronomy since pre-Greek periods. The concept of Naisargika Bala ( natural force) discussed in Hora texts from India is defined to be proportional to planetary size and inversely related to planetary distance. This idea developed several centuries prior to Isaac Newton resembles fundamental physical forces in nature especially gravity. We show that the studies on retrograde motion and Chesta Bala of planets like Mars in the context of astrology lead to development of differential calculus and planetary dynamics in ancient India. The idea of instantaneous velocity was first developed during the $1^{st}$ millennium BC and Indians could solve first order differential equations as early as $6^{th}$ cent AD. Indian contributions to astrophysics and calculus during European dark ages can be considered as a land mark in the pre-renaissance history of physical sciences.

Key words: Physical forces, differential calculus, planetary dynamics, ancient India

## 1. Introduction

It is generally believed that astrophysics and advanced mathematics is an offspring of renaissance period Europe. Several studies [1] suggest that non-European contributions especially prior to the renaissance has also played a significant role in the development of basic sciences. Indian civilization is one of the oldest of that kind in the world which has made original marks in fields like mathematics,philosophy,meteorology,metallurgy and surgery even prior to the Greek invasion period [2-6] Many Indian contributions in astronomical sciences prior to Christian era is perhaps lost which can be partly recovered from other forms of classical literature like astrology and later commentaries [7]

Development of differential calculus [8] and ideas fundamental physical forces [9] were considered to be triumphs of the European renaissance through dedicated works of Galileo,Kepler,Newton etc. In this paper In this paper we have studied the following problems

(i) The ideas of physical forces developed in ancient India and its relation with concepts in modern physics.
(ii) The development of the ideas of differential calculus and planetary dynamics in association with the ancient Indian contributions to astrophysics.

It will be shown that ancient Indian astronomers could define a planetary force which is proportional to the planetary size and inversely relatedto the planetary distance. The variation of force of interaction of planetary objects during its relative orbital motion is also studied by them using quantitative models of planetary motion during the1st millennium AD.We will also discuss the development of ideas of differential calculus in India since $1^{st}$ millennium BC. This paper summarises our earlier results based on the studies of Indian historical contributions to Physics and mathematics as a part of astronomy. [ 10-13].

## 2. Ideas of Physical Forces in ancient India

Astronomy and astrology was practiced in ancient India as an integrated discipline called Jyotisha [ 14,15]. This subject has three important branches viz.

(i) Ganita which includes theoretical and experimental astronomy with branches like karana ( computational procedures) and siddhanta ( treatises on general astronomy)

(ii) Hora Sastra Mainly principles and practice of astrology with branches like Jataka,Prasna, Muhurta,Nadi etc

(iii) Samhita which is equivalent to a general and scientific encyclopedia including astronomy, applied sciences and astrology.

Ancient Indian astronomers like Parasara,Garga and Varahamihira were known to have written texts in all the three branches of Jyotisha. Physical relations between celestial objects and terrestrial phenomena which comes under the studies on modern astrophysics is studied as a part of astrology during ancient times. Old scientific works on astrology from India will be shown to contain important scientific information related to both astrophysics and mathematics in this paper.

Parasara Hora a treatise in sanskrit written during $1^{st}$ millennium BC [16,16a] defines six types of physical forces ( shadbalas) by which Sun,moon and visible planets will physically interact with earth or among themselves. Their details are shown in Table 1 and its relation to modern physics. The scientific ideas in Naisargika and Chesta bala will further receive special attention in this paper. The original sanskrit verses used to establish our results and their sources is given in Appendix I of this paper.

## 2.1 Naisargika Bala

Naisargika Bala or natural force is an inherent property of a celestial object which possesses the following physical properties as defined in astrological texts from ancient and medieval India :-

1. This force is a constant for a given object ( eg a planet ) which does not vary with position or time [17]

2. Naisargika Bala increases in the order Saturn,Mercury,Mars,Jupiter,Venus Moon and Sun whose magnitudes are given units of Shastiamsa ( 0.5 degrees) as in Table 2 [18]

3. This force ( Fn) is defined to be proportional to the apparent size or diameter of the planetary objects concerned [18.19]

4. This force varies inversely with distance ( r) of the object from the observer ( for eg in earth) [19]

5. When we want to compare force due to more than one planetary object occupying similar celestial longitudes relative to earth ( this occurs during close planetary conjunctions which is referred as grahayudha or planetary war in ancient India) Naisargika Bala becomes an important criteria to judge to find the physically most influential planet [20-22].

If F1 and F2 are Naisargika Bala due to planets 1 and 2 situated at the similar distance r from earth ( say the planets occupy nearly the same geocentric longitude or they are in conjunction) then conceptually we have

$F1 = F(D1)/r$ ; $F2 = F(D2)/r$ according to properties 3 and 4 defined above for the Naisrgika Bala. Here D1 and D2 are the apparent diameters of the planets 1 and 2 respectively.

We can find that the ratio of the planetary Nairgika Bala as

$F1/F2 = F(D1)/F(D2)$ ...... ( 1 )

According to modern theory of Newtonian gravitation the forces due to the above planetary objects are given by

$F1 = (M1M)/r^2$ and $F2 = (M2M/r^2)$

The ratio of the gravitational forces are given by

$F1/F2 = M1/M2$ ..... ( 2 )

We have $M1 = V1d1$ and $M2 = V1d2$ here V is the volume and d is the mass density of these planets

If we assume $d1=d2$ then ,

$F1/F2 = V1/V2 = F(D1)/F(D2)$ ..... ( 3 )

So the ratio of Naisargika Balas of two planets situated at identical distances from earth as defined by ancient astronomers from India is almost identical to the ratios of the modern gravitational forces of these planets if they possess also similar mass densities [23].

**2.2 Chesta Bala and emergence of differential calculus in ancient India**

Ancient Indian astronomers observed that the apparent sizes and brightness of the planets showed temporal variations. This is particularly evident during occasions of retrograde motions of planets like Mars, Jupiter etc as seen from earth . It was also noticed that the relative speed of the planets also showed notable variations during periods of retrograde motion. We can find that these astronomical observations and subsequent studies in the context of astrology lead to the emergence of differential calculus in ancient India.

When planets like mars are in the same side of earth in their orbital motion and when it approaches closer to earth or sun then due to the differences in the eccentricity of their orbits there will be relative motion between earth and the planet. This will appear with repect to fixed stellar constellations ( zodiac signs) change from direct ( say clockwise) to retrograde motion ( say anticlockwise) of the planet from earth and then back to direct motion again. Between direct and retrograde motions there is a turning point where the planet will appear as stationary. As the planet begins the retrograde motion gradually its relative speed increases and it reaches a maxima at the centre of the retrograde loop [24,25]. See Fig 1 for the relative orbital speed changes of Mars in its closed approach to earth during the year 2003 while undergoing apparent retrograde motion.

Along with the speed the geocentric distance of the planet from earth decreases and it reaches a minima at the centre of the retrograde loop. This will appear as increase in apparent size and brightness of the planet which reaches again a maxima at the centre of the retrograde loop. See Fig 2 for these changes around August 2003 retrograde motion of Mars relative to earth.

When the geocentric longitude ($\lambda$) of the planet is found to increase with time then we observe direct motion where $(d\lambda/dt) > 0$. Retrograde motion is observed when $\lambda$ decreases with time or during $(d\lambda/dt) < 0$. Between these two types of motion we can find turning points of reversal from direct to retrograde motion or vice versa. At these turning points ( called stationary points in astronomy)  we have $(d\lambda/dt) = 0$. where  planets like Mars  will appear to be stationary for an observer in Earth.

 Ancient Indian astronomers who studied retrograde motion of planets noticed the visible physical properties of these planets  ( size or brightness) change  in conjunction with the apparent speed changes. Variations in  apparent size of the planet during retrograde motion is defined as changes in  Chesta bala or  dynamic force of the planet in ancient astrological works from India . In Parasara Hora ( 6$^{th}$ Cent BC or before) eight types of dynamical situations are defined in relation with retrograde motion of planets like Mars as seen from earth [18].The nature of the instantaneous speed and the corresponding apparent size ( Chesta bala) of the planet defined in this Hora text is given in Table 3.

Stationary points are defined as Vikala situation in ancient Hora texts ( Vikala means time variations in planetary longitude is less than one kala or one minute of arc per day) These results suggests that ancient Indian astronomers were clearly aware of the concept of instantaneous velocity as early as 6$^{th}$ cent BC. This can be considered to be the beginning of the development of differential calculus in India. From the time of Brahmagupta ( celebrated Indian astronomer and mathematician) i.e from 6$^{th}$ cent AD onwards Indians could solve simple differential equations of the type $d\lambda/dt = 0$ to find the geocentric longitudes of stationary points [26,27] These values obtained from different astronomical works from ancient/medieval  India is  comparable with the values found by the Greek astronomer Ptolemy [28].

## 3. Studies of planetary dynamics in ancient India

Observations related to spatial and temporal changes in the physical properties of planetary objects from earth played an important role in the development of planetary dynamical models in ancient India. Some of these observations are discussed in Hora texts and some are in Siddhanta texts. A synthesis of different ideas given in Indian *Jyotisha* works can provide more insight in to the history of evolution of this knowledge ( see Table. 4 )

As a planetary object moves from one zodiac sign to another as observed from earth we can observe changes in its apparent size, brightness and orbital speed. The apparent brightness of celestial object is defined in Indian astrological works as rasmi bala ( power of rays) whose values obtained from selected Indian Hora texts is given in Table 5. The geocentric longitude of observation of maxima in apparent size or brightness of a planet during its orbital motion is called Ucha and minima in the same is called Neecha .in ancient Indian astrology. Generally Ucha and Neecha points are separated by 6 signs or 180 degrees indicating the near circular nature of the planetary orbits.The planetary force on earth is maximum when the planet is situated in *uccha* and the force is minimum when the planet is situated in *neecha*.In ancient Indian Hora texts methods to interpolate planetary force ( or equivalently apparent brightness or size of the planet) at any arbitrary longitude between Ucha and Neecha is also described. Since the period of Aryabhata ( $5^{th}$ cent AD) we can find the definition of mandocca ( the geocentric longitude of minimum orbital speed) equivalent to modern apogee where the apparent size is minima. Similarly an antipodal point called nichocha is defined where the size is maxima or equivalent to modern perigee [29-32].

For an observer in earth the epoch of observation of maximum apparent size/brightness of a planetary object is during oppositions with Sun for superior planets ( Mars,Jupiter and Saturn) and during superior conjunctions for inferior planets ( Venus and Mercury). This happens during the retrograde motions of these planets and where the geocentric longitudinal separation of the planet from sun is also important. In this phenomena we have to consider the variations in the relative positions of three celestial objects viz.Sun,Earth and the planet undergoing retrograde motion ..

Since time of Aryabhata ( $5^{th}$ Cent AD) we can find epicyclic theories of planetary motion described in Indian astronomical ( Siddantic) texts. Astrological calculations, especially estimates of the magnitudes of planetary forces ( shadbalas) improved in India due to the application of epicylic theories. Methods to calculate variations in apparent diameters of planetary objects is descried in Suryasiddhanta and Siddantasiromani.
In Sripatipaddati ( a hand book of astrological calculations employing advanced mathematical models of planetary motion )methods to calculate variations in Chesta Bala [32]of planets is at any arbitrary point in its orbital motion described. This is defined proportional to the Sighra anomaly which requires use epicycle of planetary conjunction for its calculation.

Cheta Bala = 0.33 ( Sigrocca –average geocentric longitude of the planet)

The quantity in bracket is the Sighra anomaly equivalent to the angular separation of the planet from the apparent Sun.This anomaly or Chesta Bala is maximum (separation is 180) at at the centre of the retrograde loop orduring oppositions and the same is minimum during conjunctions with sun ( separation is 0). See also Fig.3. The variations in the apparent diameter of planetary objects during its apparent orbital motion and its dependence on the angular separation from the sun( during retrograde motion) is clearly mentioned in *Siddanta siromani* of Bhaskaracharya [ 30].

4.Dscussion
\
Differeremt ancient cultures contributed to the development of astrophysics as a part of their astrological and astronomical studies. Some examples are (i) Egyptian discovery of the Saros cycles of eclipses (ii) Greek classification of stars from the measurements of their apparent brightness (iii) Naked eye sunspot observations of the Chinese and (iv) Babylonian observations of the planetary phenomena including retrograde motions of planets [24,33-35]. In this context the ancient Indian contribution to astrophysics is perhaps the development of ideas of planetary physical forces discussed in this paper several centuries prior to the European renaissance.The concept of Naisargika bala is similar to Newtonian gravitation interaction as shown in section 2.1 of this paper ; an anticipation of the idea of fundamental forces in modern physics several centuries prior to the European renaissance.

Naisargika bala or natural force of planets is defined to be proportional to the size of the celestial objects and inversely related to the distance from the observer. This is first clearly mentioned in the astrological treatise Horasara by Prithuyasas during the $6^{th}$ Cent AD.There are atleast three other independent references to the planetary size dependence of the Naisargika Bala from medieval India. This idea is discussed in Kitab-ul-Hind ( 11 th cent AD) by the medieval Muslim astronomer Alberuni in his short note on the principles of Indian astrology following Laghujataka ( an abridged form of Brihatataka written by celebrated Indian astronomer Varahamihira during the $6^{th}$ cent AD) [ 36].As mentioned in Section 2 Naisargika Bala is used to compare planetary physical forces due to two or more planets occupying almost identical position in the Zodiac at a given instant of time. This phenomenon is referred as Grahayuddha or planetary war in ancient stronomical/astrological literature which happens during close planetary conjunctions in a given zodiac sign. In the medeval Indian astronomical handbook called Karanaratna written by Devacharya during the $8^{th}$ Cent AD it is explained that the planet with larger diameter (implying Naisragika bala ) will be the victor during a close planetary conjunction or Grahayuddha. A similar idea is also found in Sripatipadati.Finally Dasadyayi a well known Sanskrit commentary of the Brihatjataka of Varhamihira by the Kerala astronomer Govinda Bhattathiripad ( $13^{th}$ cent AD) also provides independent evidence of the planetary size dependence of the Naisargika Bala

One can find a definition in Suryasiddanta [30,31,37] which states that : the dynamics or quantity of motion produced by the action of a fixed force to different planetary objects is inversely related to the quantity of matter in these objects.

This definition is equivalent to the statement of Newton's second law of motion

$$a = F/M$$

It can be considered as an indirect perception of the idea of planetary mass by ancient Indian astronomers.

One of important motivations behind the development of differential calculus by different cultures seem to be scientific modelling of observed astronomical phenomena in general and retrograde motion of planets in particular. Babylonian astronomers collected several centuries of valuable observations related to retrograde motion of visible planets like Mars. The earliest scientific model in this context was developed by Indian astronomers of the Mahabharatha period like Parasara and Garga ( on or before $6^{th}$ cent BC) through the definition of Chesta bala corresponding to eight dynamical situations during the retrograde motion of a superior planet. Ancient Indian astronomers had the idea of instantaneous velocity which is known to be equivalent to the idea of derivative in modern calculus. Greek astronomers contributed towards the mathematical models of retrograde motion of planets. Following Applonius , Ptolemy could solve first order differential equations using numerical techniques like linear interpolation to find stationary points of planetary motion during $1^{st}$ cent AD.

Several Indian astronomers contributed towards the mathematical techniques of differential calculus since $5^{th}$ cent AD [ 38] Aryabhata used linear interpolation techniques to solve difference equations to prepare a Table of Sines. Brahmagupta could find derivatives of complex functions through quadratic interpolation techniques who also solved for the first time in India , differential equations to find stationary points of retrograde motion of planets . Bhaskarachrya had expertise in both physical and mathematical aspects of differential calculus which he applied to planetary dynamic models. During early renaissance period ( 14-16centuries AD) differential and integral calculus was further developed by Kerala astronomers (of now international repute) whose principal contributions is the discovery of infinite series in calculus prior to the European contributions by Gregory,Taylor and Lebniez.

We have shown in section 3 that astrophysical observations was the principal motivation behind the developments of mathematical models of planetary motion in ancient India. The physical properties of an Keplerian elliptical orbit ( ie apogee and Perigee) of a planet could be anticipated by them through their correlation with the extremas in physical properties of the planet ( maxima or minima) such as apparent size,brightness and orbital speed. The connection between astrological studies and planetary dynamics is thus a new paradigm in the history of astrophysics [39]

Indian knowledge of planetary dynamics and higher mathematics including calculus is fairly advanced prior to the beginning of the European renaissance period in the 14-$15^{th}$ century AD. After the decline of Greek and Roman Empires by $4^{th}$ cent AD the centre of scientific activity at the global level became India where the socio-political situations were stable and favorable for new knowledge creation in diversified fields.Studies in planetary dynamics in both physical and mathematical perspective by Indian astronomers in the context of astronomy as well as astrology can be considered to be important in the history sciences during the European dark ages . There are independent evidences which suggest that Arab astronomers during the medieval period ( 11[th] to 15[th] cent AD) were aware of Indian contributions planetary physical forces [40]. There is every possibility that the new knowledge created in India in areas like astrophysics and differential calculus was made available the European scientists during renaissance period directly or indirectly through Arab and Christian missionary routes of transmission .It may be worth remembering that both Isaac Newton and Kepler were aware of astrology too at the advanced level [41-42].

## Aknowledgements

The authors are grateful to Dr.S.Madhavan ( Retd Prof of Mathematics) for his interest in our work and for many useful discussions. We also express our sincere thanks to Prof.Gopalakrishnan for correcting the Sanskrit grammar and to Mrs.Sobha for her help in typing the Sanskrit verses in this paper.

**Table 1 Six types of planetary physical forces ( Shadbalas) defined ancient Hora texts from India and their equivalent concepts in modern physics.**

| The Name of Shadbala defined in Indian *Hora* Texts | Explanation | Modern Physical Idea |
|---|---|---|
| 1.Sthana Bala | $F = f(r)$ | Force change with position or distance ($r$) of the influencing object |
| 2.Kala Bala | $F = f(t)$ | Diurnal and seasonal changes in force –Time ($t$) dependent forces. |
| 3. Drik Bala | Planet-planet interactions | Many body problem |
| 4.Dig Bala | Force change with directiion | Force is a vector |
| 5.Naisargika Bala ( natural Force) | $F = kD/r$ an inherent property of the celestial obect | Force directly proportional to size (D), inversely proportional to distance ($r$) of the planets |
| 6.Chesta Bala | Dynamic force | Change in force of planets undergoing retrograde motion |

**Table 2. Ancient Indian measurements of apparent diameters of planetary objects and corresponding values of Naisargika Bala.**

| Planetary Object | Apparent diameter (Yojanas) from Aryabhateeya | Apparent diameter (arc) from Suryasiddhanta | Naisargika Bala (shastiamsas) Parasara Hora |
|---|---|---|---|
| 1. Sun | 4410 | 32' 4" | 60 |
| 2 .Moon | 315 | 31' 7" | 51.43 |
| 3 Venus | 63 | 4' | 42.85 |
| 4. Jupiter | 31.5 | 3' 30" | 34.28 |
| 5. Mercury | 21 | 3' | 25.7 |
| 6. Mars | 12.6 | 2' | 17.1 |
| 7. Saturn | 15.75 | 2' 30" | 8.57 |

**Table 3. Eight types of planetary situations during retrograde motion indicating variations of orbital speed of planet and corresponding value of Chesta Bala.**

| Dynamical situation of planet | Relative speed / nature of motion | Chesta Bala (shastiamsa) |
|---|---|---|
| 1. Anuvakra | Direct motion | 30 |
| 2. Vikala | Stationary point | 15 |
| 3. Mandatara | Very slow motion | 7.5 |
| 4. Manda | Slow motion | 15 |
| 5. Sama | Average speed | 30 |
| 6. Chara (Sighra) | Fast motion | 30 |
| 7. Ati chara (Sighratara) | Very fast motion | 45 |
| 8. Vakra | Max orbital speed/ Centre of retrograde | 60 |

**Table 4.** Extrema [maxima/minima] in different physical properties of planetary objects observed by Indian astronomers of 1st millennium BC and 1st millennium AD during the sidereal orbital motion of planets as observed from earth.

| Planetary Physical parameter | Location of maxima | Location of minima | Indian sources/remarks |
|---|---|---|---|
| 1. apparent size or brightness | Located at *Uccha* or during retrograde (*Vakra*)motion | Located at *neecha* | Indian Hora Texts like Parasara Hora (6th BC) |
| 2. orbital speed | Located at *nichocha* of the epicycle | Located at *mandocha* of epiccycle | Indian astronomical texts of 1st millennium AD. Variations in orbital speed studied in Parasara Hora |
| .3. planetary distance | at *mandocha* of epicycloe | at *nichocha* of epicycle | Indian astronomical texts of 1st millennium AD. |

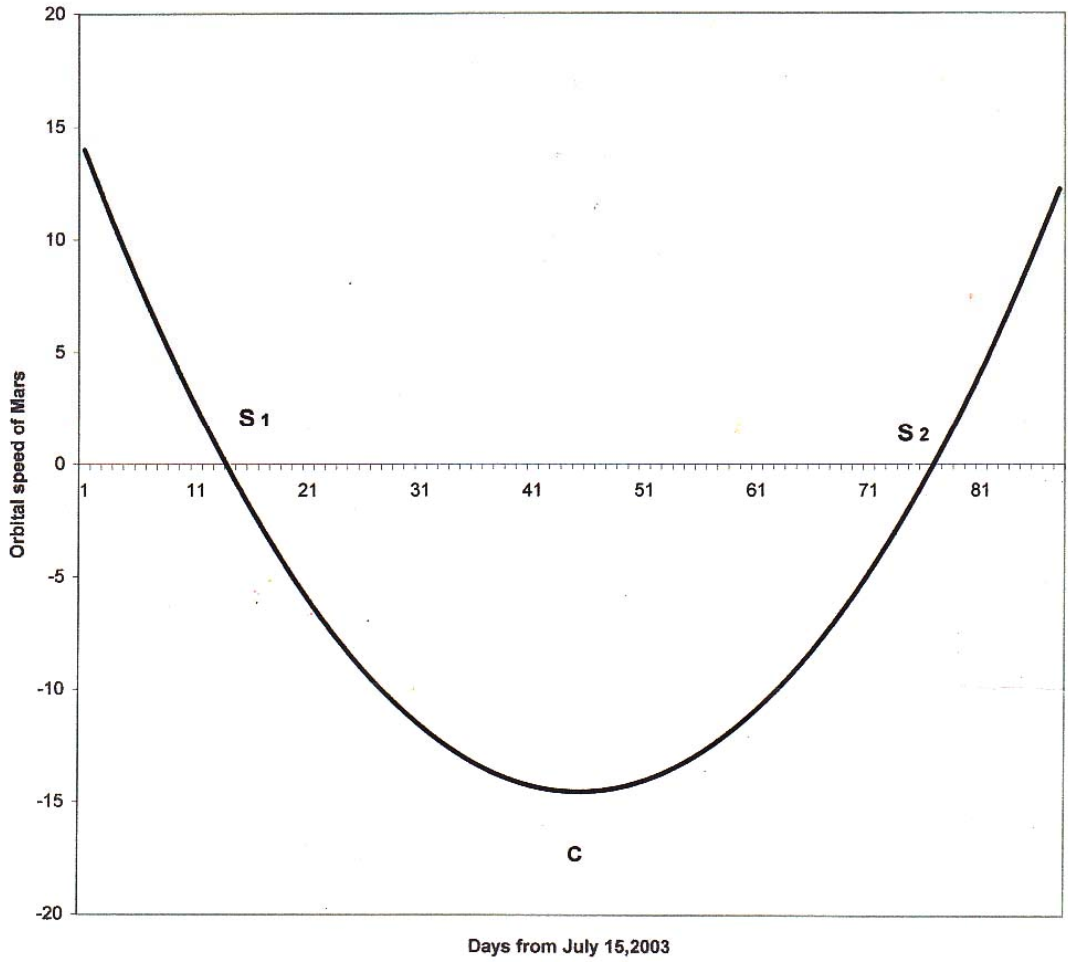

**Fig.1** Variations of the Orbital speed of Mars as observed from earth during the retrograde motion during July –October 2003. Here S1 and S2 are the stationary points

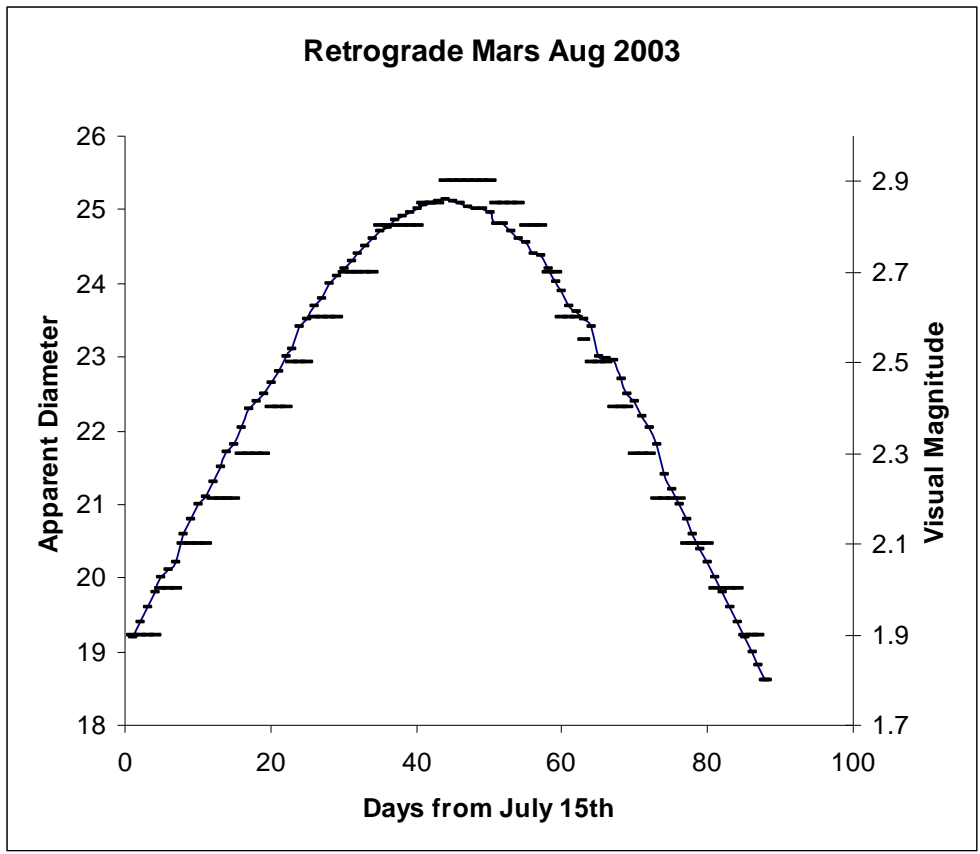

**Fig 2** Variations in the apparent diameter ( dotted) and visual magnitude ( broken) of Mars during its retrograde motion during July-October 2003. Both quantites reached a maxima at the centre of the retrograde loop during August 2003.

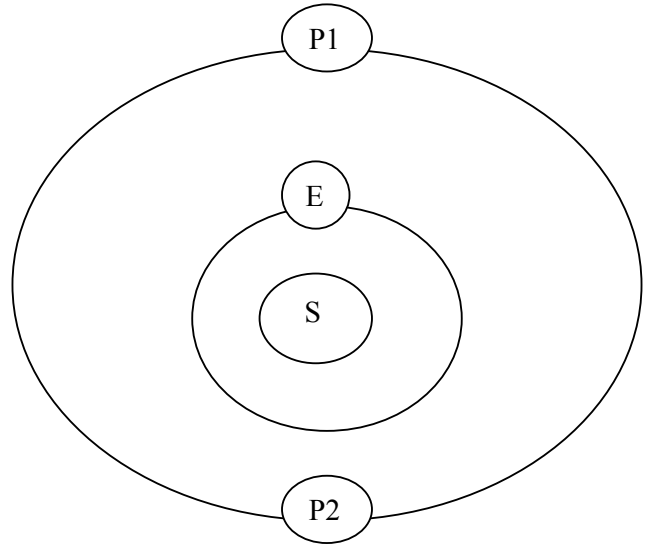

**Fig 3**. **The chesta Bala is maximum during opposition ( P1) and minimum during conjunction ( P2) for superior planet like Mars. Here E is earth, P is planet and S is sun.**

# Appendix I - Original Sanskrit verses and their Scientific meanings

1. निसर्गकालचेष्टाख्यस्थानायन दिख्याय
   षट् प्रकारः बलं प्राहुर् ग्रहाणां पूर्व सूरयः।।

   Our predecessors (ie., Indran astronomers prior to the period of Parasara) have proposed six types of planetary physical forces viz. Nisarga, Kala, Cheṣta, Sthāna, Dig and Ayana.

   Text: Parasarahōra cited in Dasādhyāyi [18] 2.21.20.

2. शरुबुगुशुचसाद्या वृद्धितो वीर्यवन्तः।

   The natural force (Naisargika Bala) of planetary objects increases in the following order: Saturn, Mars, Mercury, Jupiter, Venus, Mars and the Sun.

   Text: Bṛhat Jātaka as cited in Dasādhyāyi [18] 2.21.

3. षष्टिरेकेषवस्स दशषड् विंशतिस्ततः
   चतुस्त्रिंशत् त्रिवेदांगा सूर्यादीनाम्।
   निसर्गजा षष्टिः एकपञ्चाशत् सप्तदश
   षड्विंशतिः चतुस्त्रिंशत् त्रिचत्वारिंशत् नव
   सूर्यादीनां निसर्गबलं षष्ट्यांशः।।

   The magnitude of the natural force of different planetary objects in units of Shatiamasa (half degree of arc) is as follows: Sun (60), Moar (51), Mars (17), Mercury (26), Jupiter (34), Venus (43) and Saturn (9).

   Text: Parasara Hora cited in Dasādhyāyi [18] 2.21.52.

4. सौराद् द्विगुणो भौमो बलवान् भौमाच्चतुर्गुणः सौम्यः।
जीवोऽष्टगुणः सौम्याज्जीवादष्टाधिको भृगुर्बलवान्॥

शुक्राद्धि षोडशबलश्चन्द्रो भानुस्ततोधिको द्विगुणः।
तद्वद् द्विगुणः सर्वेषां राहुर्बलवान् निसर्गतश्च॥

The natural force (naisargika bala) of Mars of twice that of Saturn. Natural force of Mercury is twice that of Mars. The naisargika bala of Jupiter is eight times that of Mercury. The same for Venus is eight times that of Jupiter. The naisargika bala of Mars is sixteen times that of Venus. Natural force of Sun is twice that of Mars. Finally the naisargika bala of Rahu (the shadow covering earth during solar/lunar eclipses) is twice that of Sun. So compared to other celestial objects naisargika bala of Rahu is the maximum.

Text: Hōrasāra [19] 3.32 and 3.33.

5. शरबुगुशुचसाद्यः वृद्धितो वीर्यवन्तः
इति निसर्गबल कथने मण्डलवृद्धिरप्युक्तम्
तेन चन्द्रस्य ताराग्रहेभ्यो मण्डलवृद्धिरस्तीति सिद्ध्यति॥

When the naisargika bala of different celestial objects like Sun, Mars etc. is defined (in Hora texts like Bṛhat Jātaka) its association with the size of its disc (apparent diameter) is also specified. Since moon's apparent diameter is observed to be larger than that of the star planets, it is defined to have greater naisargika bala compared to them.

Text: Dasādhyāyi [18] 2.19.5.

6. भानामवस्थानगताः क्रमेण मन्दार्यभौमार्कसितज्ञचन्द्राः।
तेषामधः स्थानगतो बलीयान् राहुर्महीमण्डलमूर्धसंस्थः॥

The mean positions of different planetary objects (in their respective orbits) in the increasing order of their distance from earth is: Rahu, Moon, Mercury, Venus, Sun, Mars, Jupiter and Saturn. Since Rahu (the shadow object) is closest to earth its apparent size (equivalent ot naisargika bala) is observed to be the maximum. The result implies that naisargika bala of planets varies inversely with the distance from the observer.

Text: Hōrasāra [19] 3.35.

7.    मन्दार सौम्य वाक्पति
      सितचन्द्राक्का यथोत्तरं बलिनः।
      नैसर्गिकबलमेतद्
      बलसाम्येऽस्माद् बलाधिक चिन्ता।।

   Naisargika bala is the main criterion based on which we can compare physical forces due to more than one planet during close planetary conjunctions in a rasi at a given instant of time.

   Text: Laghujātaka [20]  2.10.

8.    शरुबुगुशुचस्याद्या नामाक्षरैः ग्रहा अवगन्तव्या
      एते 'वृद्धितो'ऽधिकमधिकं वीर्यवन्तः
      एतेषां यथोत्तरं बलवृद्धिरित्यर्थः
      बहुष्ट फलधिकग्रह निरूपणेष्ट अनेन
      प्रयोजनं नस्वाफलनिरूपणे।।

   Same as verse No. 7.

   Text: Dasādhyāyi [18] 2.21.16.

9.    उभयोरेकत्रगयोः तदा जयत्यधिकविष्कम्भः।
      उभयोरुत्तरगोले विक्षेपेणाधिकस्तदा जयति।।

   When two planets are together (during close conjunctions) the planet with greater diameter is the victor of the planetary war or grahayudda.  (Here the victor planet will exert larger naisargika bala (natural force) compared to its companion planets.)

   Text: Kāraṇa ratna [14] 8.22.

10.   अपसव्ये जितो युद्धे पिहितोऽणुरदौप्तमान्।
      रुक्षो विवर्णो विध्वस्तो विजितो दक्षिणाश्रितः।
      उद्क्स्थो दीप्तिमान् स्थूलोजयोयाम्येऽपियोबली।।

   The winner of a planetary war (two or more planets in close conjuctions) is that planet whose disc is brighter and larger compared to its companions.  This result is independent of the geocentric latitude (north - south position) of the planets in conjunction.

   Text: Sūryasiddhānta [21, 21a] 7.20 and 7.21.

11. ग्रहाणाम् अष्टधा गतिर्भवति
 षष्टिचक्रग्रते वीर्यम् अनुवक्र गतेर्ईलम्
 पादं विकलभुक्तस्याइलमेव समागमे
 पादं मन्दगतेस्तस्यदलं मन्दतरस्य च
 शीघ्रं भूक्तेस्तु पादोनं दलं शीघ्रतरस्य च इति।।

These are eight (8) types of planetary dynamical situations (during retrograde motion of planets like Mars apparent for an observor in earth). The magnitude of Chesta bala (implying planetary apparent size or brightness) during these situations in units of Shatiamsa are given by Anuvakra (30), Vikala (15), Mandatara (7.5), Manda (15), Sama (30), Sighra or Chara (30), Sighratara or Atichara (45), Vakra (60).

Text: Parasara Hōra [16] cited in Dasādhyāyi [18] 2.21.48 & 2.21.49.

12. सोच्चस्थे दश सूर्ये नव चन्द्रे पञ्च भूमितनये च
 पञ्चेन्दुंज तथेढ्चे सप्ताष्टौ भार्गव शनेः पञ्च।।

The maximum apparent brightness (rasmi bala or rasmi number) of different planetary objects (when it is located in ucca position in the 2 diac) is as follows: Sun (10), Moon (19), Mars (5), Mercury (5), Jupiter (7), Venus (8) and Saturn (5).

Text: Saravali [29] 36.2

13. चतुरेकविंशषोडशनवसंख्याः पञ्चसप्तदशकिरणाः
 मन्दगतिचन्द्रभार्गवशशीसुतरविभौमजीवानाम्।।

Rasmibala of planets given in Kṛṣṇīyam is as follows: Moon (21), Sun (5), Venus (16), Jupiter (10), Mars (7) and Saturn (4).

Text: Kṛṣṇīyam [ ] 3.4

14. नीचोनन्तु ग्रहं भार्द्धाधिके चक्राद्विशोधयेत्
    भागीकृत्य त्रिभिर्भुक्तं फलमुच्चबलं भावेत्।।

The difference between the geocentric longitude of the given planet and the longitude of its Ucca in degrees on division by three will yield Ucca bala of the planet concerned.

Text: Parasarahōra cited in Dasādhyāyi [18] 2.21.21.

15. परमसोच्चे पूर्णमकं बलं परमनीचे शून्यभित्यन्नरेनुपातः
    अशीत्युत्तरशतभागाः सोचे नीचन्तरम्।
    ततो भागत्रित यस्य एकबलः षष्ट्यंशो भवति।
    ततस्त्रीभिर्भुज्यते इति।।

In ucca position the planetary force is maximum and is equal to 60 shastiamsas. In neecha position it is minimum or zero. Between ucca and neecha we have 180 equal parts or 180 degrees of arc. One third of it is 60 (the maximum magnitude of force). So we need to divide the longitude difference (as defined in verse 14) by three.

Text: Dasādhyāyi [18] 2.21.21.

16. अभिमुखरश्मिर्न्नीचात् द्रष्टः सोचात् पराङ्मुखो ज्ञेयः
    अन्तरगतेनुपातो यथा यथा संप्रवक्ष्यामि।
    नीचविहितश्चक्रात् शुद्धः षड्भवनतो यदाभ्यधिकः
    आत्मीयरश्मिगुणिताः षड्भक्ता रश्मयस्तस्मात् इति।।

Ramibala or apparent brightness of planets is maximum in the ucca position and minimum (or zero) in the neecha position in the zodaic. When the planet moves away from neecha position its apparent brightness increases and reaches a maximum (at the ucca position). Similarly when the planet moves from ucca position its brightness decreases and becomes zero in the neecha position.

Whent he planet is locatd at any arbitrary position between ucca and neecha, the rule of three (a linear interpolation technique) is applied to calculate its rasmi bala or apparent brightness.

Text: Sārāvali [29] 36.5 and 36.6.

17. उच्चादिस्थानसंज्ञाभेदात् फलभेदाश्च सन्ति.......
    तथा च सारावल्याम्:
    सोचे तु भवति दीप्तः............

    Depending on the location of ucca the meaning and physical consequence of the given location of the planet in the zodaic will vary. In Saravali it is said that when the planet is situated in ucca it is very bright.

    Texts: (1) Dasādhyāyi [18] 2.21.7 and 2.21.8  (2) Sarāvali [29] 5.3.

18. चेष्टा बलेषु वक्रबलतुल्यमेव..........
    तथा च सारावल्याम्:
    उच्चेराशन विलोमे फलन्नान्यैर्भविष्यति
    कालस्यातिबलत्वंस्यात्तस्मात् सोचेतिवक्रिते
    वक्रिणस्तु महावीर्यः............

    When a planet is situated at the centre of the retrograde loop (Vakra position) its physical force is a maximum which is similar to situation of planets (like Sun and Moon without apparent retrograde motion) in the ucca position. In both ucca and Vakra situations planetary physical force is very high.

    Texts:  Dasādhyāyi [18] 2.21.6,  Sarāvali [29] 5.14

19. सिद्धान्तोक्त परिस्फुटोपकरणैस्ते चासकृत् कर्मणा
    भावोः खेट दृशो बलानी च ततस्तेषाम्
    विचिन्त्यानि षड् इति।

    By making the necessary astronomical observations and calculations as specified in Siddhanta texts we can determine the six types of planetary physical forces (Shadbalas) and other astronomical parameters.

    Text: Sripatipaddati, cited in  Dasādhyāyi [18] 2.21.19

20. यः स्यात्प्रदेशः प्रतिमण्डलस्य
    दूरे भुवस्तस्य कृतोच्चसञ्ज्ञा।
    सोऽपि प्रदेशश्चलतीति तस्मात्
    प्रकल्पिता तुङ्गगतिर्गतिज्ञैः॥२०॥

That point in the eccentric orbit of the planet which is at maximum distance from the Earth is defined as uccha. This point is not fixed, but moves. The neecha point in the orbit is at a distance of six signs (or equivalently $180^0$ away) from the uccha point......

Text: Siddhāntaśiromaṇi [30] Golādhyāya 5.20 and 5.21.

21. उच्चस्थितो व्योमचरः सुदूरे
नीचस्थितः स्यान्निकटे धरित्र्याः।
अतोऽणुबिंबः पृथुलश्च भाति
भानोस्तथासन्नसुदूरवर्ती।।

When the planet is situated in ucca it is at greatest distance from earth and hence its visible disc (apparent diameter) appears large. When it is situated in neecha it is at minimum or least distance from Earth and therefore its visible disc appears small. We can also observe an inverse of Variation of the apparent size of the planet with distance from the Sun (during retrograde inchain of planets like Mars).

Text: Siddhāntaśiromaṇi [30] Golādhyāya 5.22.

22. महत्त्वात्मण्डलस्यार्कः स्वल्पमेवापकृष्यते।
मण्डलाल्पतया चन्द्रस्ततो बह्वपकृष्यते।।

भौमादयोऽल्पमूर्तित्वाच्छीघ्रमन्दोच्चसंज्ञकैः।
देवतैरपकृष्यन्ते सुदूरमतिवेगिताः।।

अतो धनर्गो सुमहत् तेषां गतिवशाङ्गवेत्।
च्चाकृष्यमाणस्तैरेवंव्योम्नियान्त्यनिलाहताः।।

The quantity of motion Mhar produced in Sun due to the action of an attractive force is very small compared to Moon because of its large physical size or dimensions compared to Moon.

Since the physical size of visible planets in the solar system like Mars is very small compared to the Sun and Moon the effect of the forces on them will be very large. Thus the displacements (positive or negative from reference points) produced in them will be also of a larger magnitude due to the result of action of these forces.

Text: Sūryasiddhānta [21, 21a] 2.9, 2.10 and 2.11.